\documentclass[runningheads]{llncs}
\usepackage[T1]{fontenc}
\usepackage{graphicx}
\usepackage{amsmath}
\usepackage{subcaption}
\usepackage{booktabs}
\usepackage{multirow}
\usepackage{amssymb}
\usepackage[misc]{ifsym}
\begin{document}
\title{NGAT: A Node-level Graph Attention Network for Long-term Stock Prediction}

\titlerunning{Node-level Graph Attention Network}

\author{Yingjie Niu\Letter\inst{1,2}\orcidID{0000-0001-9322-2726} \and
Mingchuan Zhao\inst{3}\orcidID{0000-0003-0871-3614} \and
Valerio Poti\inst{4}\orcidID{0000-0003-1156-5616} \and
Ruihai Dong\inst{1,2}\orcidID{0000-0002-2509-1370}}
%
\institute{School of Computer Science, University College Dublin, Dublin, Ireland \and
SFI Centre for Research Training in Machine Learning, Dublin, Ireland \and
School of Mathematical Sciences, Dublin City University, Dublin, Ireland \and
Michael Smurfit Business School, University College Dublin, Dublin Ireland \\
\email{yingjie.niu@ucdconnect.ie, mingchuan.zhao@dcu.ie, \{valerio.poti,ruihai.dong\}@ucd.ie}}

\maketitle              
\begin{abstract}
Graph representation learning methods have been widely adopted in financial applications to enhance company representations by leveraging inter-firm relationships. However, current approaches face three key challenges: (1) The advantages of relational information are obscured by limitations in downstream task designs; (2) Existing graph models specifically designed for stock prediction often suffer from excessive complexity and poor generalization; (3) Experience-based construction of corporate relationship graphs lacks effective comparison of different graph structures. To address these limitations, we propose a long-term stock prediction task and develop a Node-level Graph Attention Network (NGAT) specifically tailored for corporate relationship graphs. Furthermore, we experimentally demonstrate the limitations of existing graph comparison methods based on model downstream task performance. Experimental results across two datasets consistently demonstrate the effectiveness of our proposed task and model. The project is publicly available on GitHub to encourage reproducibility and future research.\footnote{\url{https://github.com/FreddieNIU/NGAT}}

\end{abstract}
\section{Introduction}
Graph Neural Networks (GNNs) have become the dominant approach for graph representation learning due to their message-passing mechanisms, which aggregate neighboring nodes to learn contextual embeddings reflecting graph structures. This success has spurred interest in applying graph representation learning to stock prediction tasks, where companies are modeled as nodes with inter-firm relationships as edges \cite{li2021modeling,xu2022hgnn} to capture momentum spillover effects. Existing studies typically follow a common paradigm: 1) modeling corporate relationships to construct edges, 2) applying GNNs to Corporate Relationship Graphs (CRGs) for node embedding aggregation, and 3) evaluating performance through downstream tasks. However, this paradigm presents three key challenges:

First, momentum spillover often exhibits lagged effects, where historical trends of one stock influence another's future trends over uncertain periods (potentially exceeding one day). Existing graph-based stock prediction studies predominantly focus on next-day trend prediction \cite{xu-cohen-2018-stock,soun2022accurate} with single-day prediction horizons, which may fail to fully exploit graph models' capabilities. Furthermore, most models emphasize stock returns while neglecting volatility—a critical financial risk indicator essential for robust market analysis. To bridge these gaps, we introduce a long-term stock return/volatility prediction task that forecasts average returns or volatility over future periods using historical data and relational graphs. This formulation better aligns with momentum spillover characteristics and provides actionable insights for long-term investors.

Second, general GNN architectures like Graph Convolutional Networks (GCNs) and Graph Attention Networks (GATs), despite their generalization capabilities, may not directly suit CRGs. In conventional homogeneous graphs (e.g., Cora dataset \cite{bojchevski2017deep}), nodes share similar feature distributions within categories, making shared attention mechanisms effective. However, in CRGs where each node represents a distinct company with unique stock behavior patterns, assuming identical distributions across nodes becomes untenable. While specialized architectures for stock prediction exist \cite{feng2019temporal,cheng2021modeling}, their complexity often compromises generalizability. To address this, we propose a Node-level Graph Attention Network (NGAT) that assigns each company a unique attention mechanism, enabling tailored modeling of CRGs. Crucially, NGAT operates as a portable graph convolution layer adaptable to diverse graph-based tasks and datasets.

Third, the construction of CRGs predominantly relies on experience driven approaches, with no consensus on optimal methodologies. We compare the performance of identical GNN models under different graph constructions, demonstrating the limitations of relying solely on downstream task performance for graph quality assessment.

Our main contributions are summarized as follows:
\begin{itemize}
\item Designed a novel long-term stock prediction task that better captures momentum spillover effects;
\item Developed a highly generalizable Node-level Graph Attention Network (NGAT) that better suits corporate relationship graphs (CRGs);
\item Demonstrated empirically the limitations of model-based graph comparison;
\item Validated by experimental results the practical utility of our task and demonstrate NGAT's superior performance over baselines across multiple tasks.
\end{itemize}

\section{Related Work}

Existing research on stock prediction evolves along two key dimensions: (1) task formulation progressing from traditional econometric modelling to machine learning paradigms, and (2) architectural innovation transitioning from generic models to financial-specific adaptations. 

The foundation of stock prediction rests on econometric models including factor analysis \cite{fama2015five,carhart1997persistence}, time-series approaches \cite{poterba1988mean,lo1990when}, and volatility modelling via GARCH variants \cite{engle1993measuring,nelson1991conditional} and HAR-RV \cite{corsi2009simple}. Machine learning advancements introduced hybrid models combining GARCH with neural networks \cite{kim2018forecasting,persio2021forward} and sequence models like LSTM \cite{selvin2017stock,sunny2020deep,jin2020stock}. While these methods initially focused on next-day prediction \cite{xu-cohen-2018-stock}, recent works explore ranking tasks \cite{feng2019temporal}, overnight trend prediction \cite{li2021modeling}, and price-limit hitting forecasting \cite{xu2022hgnn}. Our proposed long-term return/volatility prediction task overcomes the limitation that next-day prediction cannot fully exploit the benefit of learning momentum spillovers by jointly modeling multi-period spillovers and risk factors through corporate relationship graphs.

The integration of alternative data sources—particularly textual information \cite{bollen2011twitter,atkins2018}—motivated graph-based approaches using GCN \cite{kipf2016semi} and GAT \cite{velivckovic2017graph} to model inter-firm relationships via price correlations \cite{li2021modeling}, industry taxonomies \cite{feng2019temporal}, or news co-occurrences \cite{niu2024evaluating}. Subsequent domain-specific adaptations introduced temporal-GNNs \cite{xiang2022temporal}, AD-GAT \cite{cheng2021modeling}, and hierarchical GNN architectures \cite{xu2022hgnn}, but at the cost of increased complexity and reduced generalizability. 
Our NGAT architecture resolves this trade-off through node-specific attention mechanisms while enabling portable deployment. 

\section{Problem Formulation} \label{sec:problem formulation}
To overcome the limitations of the next-day trend prediction task, we introduce the long-term return mean/volatility prediction tasks in this paper. The objectives of these tasks are to forecast the stock return trend and volatility over the future $T$ business days. The return trend prediction is framed as a classification task, aiming to predict whether the average return over the upcoming period will be higher or lower than in the previous period. The daily log return of stock $s$ on day $d$ is calculated as: $r_d^s = log(p_d^s / p_{d-1}^s)$ where $p_d^s$ denotes the adjusted closing price of stock $s$ at day $d$. The classification label $y_{d,c}^s$ is calculated as:
\begin{equation} \label{eq:mean_task}
    y_{d,c}^s =  \textbf{1} ( \frac{1}{T} \sum_{i=d+1}^{d+T} r_i^s > \frac{1}{T} \sum_{i=d-T+1}^d r_i^s ),
\end{equation}
where $T$ is the fixed period length. This task allows for the categorization of return trends into the following four scenarios: 
\begin{itemize}
    \item $(\mathrm{LP}, \mathrm{N+})$: \textbf{L}ast period \textbf{P}ositive return (LP) followed by an even \textbf{higher} return in the next period (N+), indicating a "surge."
    \item $(\mathrm{LN}, \mathrm{N+})$: \textbf{L}ast period \textbf{N}egative return (LN) followed by a \textbf{higher} return in the next period (N+), indicating a "rebound."
    \item $(\mathrm{LP}, \mathrm{N-})$: \textbf{L}ast period \textbf{P}ositive return (LP) followed by a \textbf{lower} return in the next period (N-), indicating a "pullback."
    \item $(\mathrm{LN}, \mathrm{N-})$: \textbf{L}ast period \textbf{N}egative return (LN) followed by an even \textbf{lower} return in the next period (N-), indicating a "plunge."
\end{itemize}

On the other hand, volatility prediction is treated as a regression task, focused on forecasting the exact value of the volatility over the upcoming period. In our experiment, we use the realized sample standard deviation to estimate the volatility, which is calculated as follows:
\begin{equation}
    y_{d,r}^s = \sqrt{\frac{1}{T} \sum_{i=d+1}^{d+T} (r_i^s - \mu)^2}, \quad \mu = \frac{1}{T} \sum_{i=d+1}^{d+T} r_i^s,
\end{equation}
where $\mu$ is the mean of log returns in the period $[d+1, d+T]$. 


The $\mathrm{N+}$ scenarios suggest market opportunities, while $\mathrm{N-}$ scenarios signal downside risks. Specifically, identifying $(\mathrm{LP}, \mathrm{N+})$ points to a likely surge, recommending a "hold" or "increase holdings" strategy. Conversely, $(\mathrm{LN}, \mathrm{N-})$ serves as a warning for potential loss, advising "sell" or "short sell" strategies. By understanding these scenarios, investors can more accurately assess future stock price trends. Furthermore, integrating predictions of long-term trends with volatility forecasts enables a more robust evaluation of the reliability and associated risks of purported "opportunities". Together, these predictions provide a more comprehensive understanding of market dynamics, thereby enhancing long-term investment strategy formulation.

\section{Model Architecture}
In this section, we describe the proposed model architecture, including the sequential embedding module, relational embedding module, and output layer. 

\subsection{Sequential Embedding}
To capture the temporal dependencies in stock movements, researchers commonly encode historical trading data using RNNs to generate sequential embeddings for stock prediction \cite{li2016tensor,chen2018incorporating,li2020multimodal,feng2019temporal,cho2014learning}. Following the common practice, LSTM is selected as the sequential embedding module in this paper. Given a set of assets $\mathbf{S}$, to generate the sequential embedding for stock $s \in \mathbf{S}$ on day $d$, we fed a time-series feature matrix $X_d^s$ into the LSTM module.

\begin{equation}
    h_d^s = \textit{LSTM}(X_d^s),
\end{equation}
where $h_d^s$ is the sequential embedding of stock $s$ on day $d$ and $X \in \mathbb{R}^{\Delta d \times F}$. $\Delta d$ is a fixed lag size, and $F$ is the number of features. We use the output of the last time step as the sequential embeddings.

\begin{figure*}
    \centering
    \includegraphics[width=\linewidth]{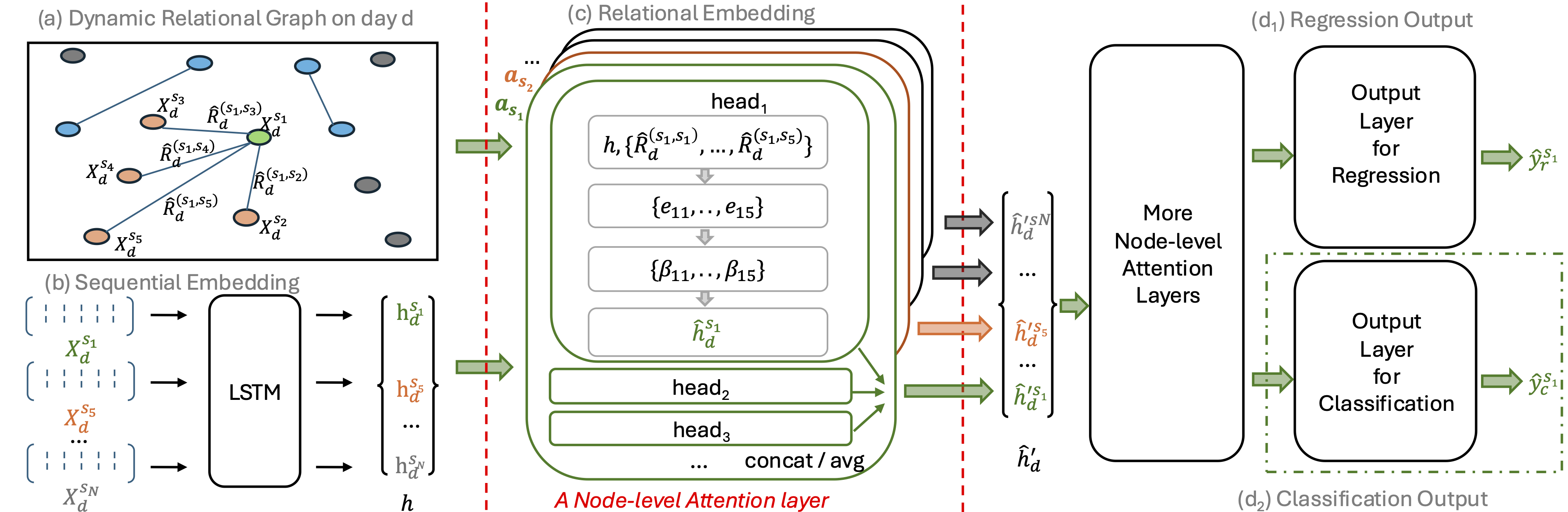}
    \caption{Node-level Graph Attention Network model architecture and calculating pipeline.}
    \label{fig:model-architecture}
\end{figure*}

\subsection{Relational Embedding} \label{sec:relational_embedding}
Beyond temporal dependencies on historical transaction data, a stock's trend is influenced by related stocks. Thus, we fuse the sequential embeddings using stock relationships to generate the corresponding relational embedding.

\textbf{Relation Building}  
Edges are constructed based on firm co-occurrences in news/tweets. For each article $n_d^k$ (the $k$-th news on day $d$), we extract its mentioned firm set $S$, then assign $R_d^{(s_i,s_j),k} = 1$ for all $(s_i,s_j) \in S \times S$. Daily aggregation yields unconditional relation attributes: $R_d^{(s_i, s_j)} = \sum_{k=1}^K R_d^{(s_i, s_j), k}$ where $K$ is the daily news count. Each node receives a self-loop with the attribute: $R_d^{(s_i,s_i)} = \max_{j \in \mathbb{N}_i} R_d^{(s_i,s_j)}$ where $\mathbb{N}_i$ is the neighbors of $s_i$. To model persistent news impacts, we compute \textit{conditional relations} via temporal weighted aggregation:  
\begin{equation}
\hat{R}_d^{(s_i,s_j)} = \sum_{m=0}^{\delta-1} w_m \cdot R_{d-m}^{(s_i,s_j)}, \quad w_m = \frac{\delta - m}{\sum_{n=1}^{\delta} n},
\end{equation}
where $\delta$ is the memory window size, and $w_m$ is the aggregated weight to apply on the unconditional attribute on the day $d-m$, implementing a forgetting mechanism that prioritizes recent days.

\textbf{Node-level Graph Attention Layer} 
As illustrated in Fig.~\ref{fig:model-architecture}, our NGAT layer transforms inputs (the sequential embeddings) to outputs through the following steps. Let $\mathbf{h} = \{h_d^1, ..., h_d^N\}$ ($h_d^i \in \mathbb{R}^F$) denote the sequential embeddings of $N$ nodes, with time subscript $d$ omitted when unambiguous. First, each node undergoes three linear projections parameterized by ${\mathbf{W}_q}, \mathbf{W}_k, \mathbf{W}_v \in \mathbb{R}^{F' \times F}$. We then compute \textit{masked attention} coefficients for connected nodes:  
\begin{equation}
e_{ij} = \hat{R}_d^{(i,j)} \cdot a_i(\mathbf{W}_q h^i, \mathbf{W}_k h^j), \quad j \in \mathbb{N}_i,
\label{eq:eij}
\end{equation}  
where $a_i$ denotes node $i$'s attention mechanism and $\hat{R}_d^{(i,j)}$ the conditional relation attribute. We normalize attention weights via:  
\begin{equation}
\beta_{ij} = \mathrm{Softmax}_j(e_{ij}) = \frac{\exp(e_{ij})}{\sum_{m \in \mathbb{N}_i} \exp(e_{im})}.
\label{eq:alphaij}
\end{equation}
The attention mechanism $a$ employs node-specific parameters $\mathbf{a}_i \!\in\! \mathbb{R}^{2F' \times F'}$ (aggregated as $\mathbf{a} \!\in\! \mathbb{R}^{N \times 2F' \times F'}$), enabling unique computation per node. Incorporating equations~\ref{eq:eij}-\ref{eq:alphaij} with LeakyReLU (slope=0.2), the final coefficients derive from:  
\begin{equation}
    \resizebox{0.888\linewidth}{!}{$\beta_{ij} = \frac{\mathrm{exp}(\hat{R}_d^{(i, j)}\textit{LeakeyReLU}(\mathbf{a}_i^T[\mathbf{W}_qh^i||\mathbf{W}_kh^j]))}{\sum_{m \in \mathbb{N}_i}\mathrm{exp}(\hat{R}_d^{(i, m)}\textit{LeakeyReLU}(\mathbf{a}_i^T[\mathbf{W}_qh^i||\mathbf{W}_kh^m]))}$},
\end{equation}
where $.^T$ represents transposition, $||$ represents the concatenation operation, and $\mathbf{a}_i \in \mathbb{R}^{2F' \times F'}$ represents the attention matrix corresponding to node $i$. 

The attention coefficients generate \textit{relational embeddings} through neighborhood aggregation. While the Softmax normalization (Eq.~\ref{eq:alphaij}) ensures comparable coefficients, its smoothing effect may cause over-smoothing by over-emphasizing neighborhood information. To preserve node-specific patterns, we concatenate the transformed sequential embedding $\mathbf{W}_v h^i$ with the relational embedding $\sum_{j \in \mathbb{N}_i} \beta_{ij} \mathbf{W}_v h^j$, followed by a nonlinear projection:  
\begin{equation}
\hat{h}^i = \sigma\left(\mathbf{W} \left[\mathbf{W}_v h^i \, \Vert \, \sum_{j \in \mathbb{N}_i} \beta_{ij} \mathbf{W}_v h^j \right]\right), \quad \mathbf{W} \in \mathbb{R}^{F' \times 2F'},
\end{equation}  
where $\Vert$ denotes concatenation. Following \cite{velivckovic2017graph}, multi-head attention ($M$ independent heads) stabilizes learning via concatenated/averaged outputs. If concatenate:
\begin{equation}
    \hat{h}'^i = \Vert_{m=1}^M \sigma(\mathbf{W^m}[\mathbf{W}_v^mh^i||\sum_{j \in \mathbb{N}_i}\beta_{ij}^m\mathbf{W}_v^mh^j]),
\end{equation}
where $\beta_{ij}^m$ is the attention coefficient of the $m$-th attention head. If average:
\begin{equation}
    \hat{h}'^i = \sigma(\frac{1}{M}\sum_{m=1}^M \mathbf{W^m}[\mathbf{W}_v^mh^i||\sum_{j \in \mathbb{N}_i}\beta_{ij}^m\mathbf{W}_v^mh^j]).
\end{equation}

\subsection{Output Layer}
The output layer parameterized by a weight matrix $\mathbf{W}_o$ is implemented as follows: 
\begin{equation}
\hat{y}^i = \begin{cases} 
\mathbf{W}_o \hat{h}'^i & \text{(regression)} \\
\text{Sigmoid}(\mathbf{W}_o \hat{h}'^i) & \text{(classification)}.
\end{cases}
\end{equation}  

\textbf{Loss Function} For regression task, the Mean Square Error loss between $y_r^i$ and $\hat{y}_r^i$ is used to back-propagate and learn the model parameters. The Binary Cross Entropy loss between $y_c^i$ and $\hat{y}_c^i$ is used instead for the classification task.

\section{Experiments}

In this section, we detail our datasets, training setup, baseline models, and evaluation metrics. 

\subsection{Data}
\label{sec:numerical_data}

We use two publicly available daily based datasets. The SPNews Dataset \cite{niu2024evaluating} includes news and transaction data for S\&P500 companies, with daily bulletins listing the latest eight news items per company. To construct accurate relational graphs, redundant repetitions across days and companies are removed, ensuring each news item is recorded once for all involved companies. The ACL2018 Dataset \cite{xu-cohen-2018-stock} contains tweets and transaction data spanning two years. Text data is derived from Twitter posts, where edges between companies are defined by the co-occurrence of ticker tags (e.g. $\#$aapl, $\#$goog). Repeated mentions of the same company within a tweet, often used for emphasis, are removed to ensure accurate co-occurrence measurement.

The numerical data in both datasets include daily trading information for the companies involved, such as open ($O$), high ($H$), low ($L$), close ($C$), adjusted close ($A$), and volume ($V$). A three-step regularization process is implemented: (a) Open prices are standardized via $O' = \frac{O - \mathbb{E}(O)}{\sigma(O)}$, where $\mathbb{E(\cdot)}$ and $\mathbb{\sigma(\cdot)}$ denote the mean and standard deviation, calculated individually for each stock. (b) High, low, and close prices are expressed as rates of change relative to open prices: $H' = \frac{H - O}{O}$, $L' = \frac{L - O}{O}$, $C' = \frac{C - O}{O}$. (c) Adjusted close prices are converted into stock returns. This process standardizes prices across stocks and reduces information redundancy due to collinearity between open prices and other variables.

\subsection{Training Setup}
All parameters are selected via grid search on validation performance:  
\textit{Graph Construction}: Feature dimensions $F \!\in\! [8,12]$; memory window $\delta \!\in\! [1,2,3,5,7,10]$ (optimal $\delta\!=\!5$); relation threshold $\in [0,1]$; forecasting horizon $T \!\in\! [1,5,10,21]$ (business days: 1d/1w/2w/1mo).  
\textit{Model Architecture}: Attention heads $\!\in\! [1,4,8]$; hidden size $F' \!\in\! [16,32,64]$; dropout $\!\in\! [0,0.2,0.4,0.6]$; fixed graph layers $=1$.  
\textit{Training}: Adam optimizer ($\mathit{lr}\!=\!10^{-4}$, weight decay $5\!\times\!10^{-4}$) with Glorot initialization \cite{kingma2014adam,glorot2010understanding}. Fixed estimation window $\Delta d\!=\!21$.

\subsection{Baselines}
We compare the proposed NGAT model with both non-graph-based models and graph-based models. 

\textbf{LSTM:} A single-layer Long Short-term Memory (LSTM) model. 

\textbf{GCN \cite{kipf2016semi}:} with one sequential embedding layer and one convolutional layer. 

\textbf{GAT \cite{velivckovic2017graph}:} with one sequential embedding layer and self-attention layer. 

\textbf{LSTM+GCN \cite{chen2018incorporating}:} The results of GCN and LSTM are concatenated for prediction. 

\textbf{TGC \cite{feng2019temporal}:} A Temporal Graph Convolutional Network (TGC).  

\textbf{AD-GAT \cite{cheng2021modeling}:} Attribute-Driven Graph Attention Network captures the attribute-sensitive momentum spillovers of the neighbour nodes. We use the implementation of AD-GAT from the original paper.

\subsection{Evaluation Metrics}

For classification models predicting return trends, we use overall accuracy (ACC), the Matthews Correlation Coefficient (MCC), and the area under the ROC curve (AUC). ACC and AUC are commonly used in classification tasks, while MCC provides a fair evaluation when class distributions are imbalanced. Given the confusion matrix $\begin{pmatrix} TP & TN \\ FP & FN \end{pmatrix}$, where $TP$, $TN$, $FP$, and $FN$ represent true positives, true negatives, false positives, and false negatives, respectively, MCC is computed as:

\begin{equation}
    \mathrm{MCC} = \frac{TP \times TN - FP \times FN}{\sqrt{(TP + FP)(TP + FN)(TN + FP)(TN + FN))}}.
\end{equation}

For regression models predicting future volatility, we use out-of-sample $R^2$ and Mean Squared Error (MSE), both of which are commonly used evaluation metrics. To ensure comprehensive evaluation, we compute the average metrics across all companies in each dataset: 76 companies in ACL2018 and 268 companies in SPNews.

\section{Results and Discussion}

\textbf{Return Trend} Table~\ref{tab:r-mean} demonstrates NGAT's superior performance in return mean classification across both datasets. Notably, NGAT achieves significant improvements over GAT, with relative gains of $4.03\%$, $5.90\%$, and $2.07\%$ for $T=21$ on ACL2018, and $2.08\%$, $8.12\%$, and $1.41\%$ for $T=10$ on SPNews. These results validate the universal superiority of our node-level attention mechanism in financial modeling. Crucially, all models exceed $70\%$ accuracy – substantially outperforming traditional next-day prediction benchmarks (typically $50-60\%$) – confirming the practical value of long-term forecasting. Additionally, high MCC and AUC values indicate balanced performance across classes, further validating effectiveness of prediction. The underperformance of AD-GAT stems from its exclusion of textual embeddings, unlike its original design.

Our four market scenarios (Sec.~\ref{sec:problem formulation}) reveal strategic insights: While NGAT shows slightly weaker "rebound" $\mathrm{(LN,N+)}$ prediction on ACL2018, it excels in identifying "surge" $\mathrm{(LP,N+)}$ and "plunge" $\mathrm{(LN,N-)}$ opportunities – critical for crisis avoidance and profit capture. On SPNews, NGAT achieves balanced performance with $3.2\%$ $\mathrm{N+}$ accuracy over LSTM, despite minor $\mathrm{N-}$ tradeoffs, demonstrating robust pattern recognition across market phases.

\begin{table*}[]
\caption{Return Mean Prediction Performance (Five-run average). Notations: \textit{r-mean\textsubscript{X}} denotes T=X classification task (e.g., \textit{r-mean1}: T=1).}
\label{tab:r-mean}
\renewcommand{\arraystretch}{0.95} 
\resizebox{\textwidth}{!}{\begin{tabular}{llrrrrrrrrrrrr}\toprule
\multirow{2}{*}{\textbf{Dataset}} & \multicolumn{1}{c}{\multirow{2}{*}{\textbf{Model}}} & \multicolumn{3}{c}{\textbf{r-mean1}} & \multicolumn{3}{c}{\textbf{r-mean5}} & \multicolumn{3}{c}{\textbf{r-mean10}} & \multicolumn{3}{c}{\textbf{r-mean21}} \\ \cmidrule(lr){3-5}\cmidrule(lr){6-8}\cmidrule(lr){9-11}\cmidrule(lr){12-14}
& \multicolumn{1}{c}{} & \multicolumn{1}{c}{ACC} & \multicolumn{1}{c}{MCC} & \multicolumn{1}{c}{AUC} & \multicolumn{1}{c}{ACC} & \multicolumn{1}{c}{MCC} & \multicolumn{1}{c}{AUC} & \multicolumn{1}{c}{ACC} & \multicolumn{1}{c}{MCC} & \multicolumn{1}{c}{AUC} & \multicolumn{1}{c}{ACC} & \multicolumn{1}{c}{MCC} & \multicolumn{1}{c}{AUC} \\ \midrule
\multirow{6}{*}{ACL2018} & LSTM & 0.7345 & 0.3734 & 0.8291 & 0.7445 & 0.3763 & 0.7933 & 0.7358 & 0.3271 & 0.7643 & 0.6963 & 0.3609 & 0.7994 \\
& GCN & 0.7211 & 0.3184 & 0.8018 & 0.7466 & 0.3447 & 0.7734 & 0.7516 & 0.2992 & 0.7415 & 0.7045 & 0.3289 & 0.7510 \\
& GAT & 0.7361 & 0.3792 & 0.8357 & 0.7611 & 0.4095 & 0.8111 & 0.7521 & 0.3586 & 0.7822 & 0.6789 & 0.3579 & 0.7918 \\
& LSTM+GCN & 0.7379 & \textbf{0.3852} & 0.8317 & 0.7555 & 0.3928 & 0.7986 & 0.7476 & 0.3465 & 0.7648 & 0.7000 & 0.3727 & 0.8030 \\
& TGC & 0.7018 & 0.2788 & 0.7806 & 0.7550 & 0.3480 & 0.7786 & 0.7513 & 0.3134 & 0.7486 & 0.6971 & \textbf{0.3846} & 0.7992 \\
& AD-GAT & 0.6571 & 0.2105 & 0.7147 & 0.7313 & 0.2996 & 0.7497 & 0.7211 & 0.2757 & 0.7046 & 0.6553 & 0.2921 & 0.7375 \\
& NGAT (Our) & \textbf{0.7395} & 0.3825 & \textbf{0.8370} & \textbf{0.7774} & \textbf{0.4150} & \textbf{0.8172} & \textbf{0.7745} & \textbf{0.3652} & \textbf{0.7849} & \textbf{0.7063} & 0.3790 & \textbf{0.8082} \\ \midrule
\multirow{6}{*}{SPNews} & LSTM & 0.7569 & 0.4754 & 0.8345 & 0.7449 & 0.4643 & 0.8263 & 0.7195 & 0.4098 & 0.8151 & 0.7287 & 0.3920 & 0.7999 \\
& GCN & 0.7571 & 0.4738 & 0.8341 & 0.7418 & 0.4594 & 0.8232 & 0.7230 & 0.4175 & 0.8161 & 0.7348 & 0.4126 & 0.8083 \\
& GAT & 0.7582 & 0.4764 & 0.8388 & 0.7421 & 0.4589 & 0.8256 & 0.7233 & 0.4127 & 0.8210 & 0.7336 & 0.4039 & 0.8053 \\
& LSTM+GCN & 0.7589 & 0.4797 & 0.8382 & 0.7381 & 0.4535 & 0.8219 & 0.7210 & 0.4078 & 0.8147 & 0.7374 & 0.4162 & 0.8016 \\
& TGC & 0.7593 & 0.4727 & 0.8391 & 0.7363 & 0.4500 & 0.8220 & 0.7221 & 0.4137 & 0.8193 & 0.7358 & 0.3921 & 0.8015 \\
& AD-GAT & 0.6861 & 0.3003 & 0.7397 & 0.6870 & 0.3485 & 0.7676 & 0.6911 & 0.3621 & 0.7886 & 0.7151 & 0.3421 & 0.7537 \\
& NGAT (Our) & \textbf{0.7627} & \textbf{0.4835} & \textbf{0.8407} & \textbf{0.7504} & \textbf{0.4711} & \textbf{0.8296} & \textbf{0.7384} & \textbf{0.4462} & \textbf{0.8326} & \textbf{0.7397} & \textbf{0.4168} & \textbf{0.8142} \\ \bottomrule
\end{tabular}}
\end{table*}

\begin{table*}[t]
\caption{Accuracy of predicting return trends in different scenario groups.}
\label{tab:r-mean_group}
\centering
\resizebox{\textwidth}{!}{
\begin{tabular}{lllrrlrrlrr}
\toprule
\multicolumn{1}{c}{\textbf{Dataset}} & \multicolumn{1}{c}{\textbf{Model}} & \multicolumn{3}{c}{\textbf{r-mean5}} & \multicolumn{3}{c}{\textbf{r-mean10}} & \multicolumn{3}{c}{\textbf{r-mean21}} \\ \cmidrule(lr){3-5} \cmidrule(lr){6-8} \cmidrule(lr){9-11} 
\multirow{6}{*}{ACL2018} & \multirow{3}{*}{LSTM} & & \multicolumn{1}{c}{\textbf{LN}} & \multicolumn{1}{c}{\textbf{LP}} & & \multicolumn{1}{c}{\textbf{LN}} & \multicolumn{1}{c}{\textbf{LP}} & & \multicolumn{1}{c}{\textbf{LN}} & \multicolumn{1}{c}{\textbf{LP}} \\ 
 & & \textbf{N+} & \multicolumn{1}{r}{611/801 (0.7627)} & 719/893 (0.8051) & \textbf{N+} & \multicolumn{1}{r}{487/630 (0.7730)} & 734/911 (0.8057) & \textbf{N+} & \multicolumn{1}{r}{369/467 (0.7901)} & 495/633 (0.7819) \\
 & & \textbf{N-} & \multicolumn{1}{r}{764/1039 (0.7353)} & 735/1067 (0.6884) & \textbf{N-} & \multicolumn{1}{r}{724/996 (0.7269)} & 851/1263 (0.6737) & \textbf{N-} & \multicolumn{1}{r}{713/1027 (0.6942)} & 1069/1673 (0.6389) \\ \cmidrule(lr){2-11} 
 & \multirow{3}{*}{NGAT} & & \multicolumn{1}{c}{\textbf{LN}} & \multicolumn{1}{c}{\textbf{LP}} & & \multicolumn{1}{c}{\textbf{LN}} & \multicolumn{1}{c}{\textbf{LP}} & & \multicolumn{1}{c}{\textbf{LN}} & \multicolumn{1}{c}{\textbf{LP}} \\ 
 & & \textbf{N+} & \multicolumn{1}{r}{622/801 (0.7765)} & 731/893 (0.8186) & \textbf{N+} & \multicolumn{1}{r}{465/630 (0.7380)} & 750/911 (0.8232) & \textbf{N+} & \multicolumn{1}{r}{355/467 (0.7601)} & 491/633 (0.7756) \\ 
 & & \textbf{N-} & \multicolumn{1}{r}{831/1039 (0.7998)} & 770/1067 (0.7216) & \textbf{N-} & \multicolumn{1}{r}{794/996 (0.7971)} & 934/1263 (0.7395) & \textbf{N-} & \multicolumn{1}{r}{711/1027 (0.6923)} & 1091/1673 (0.6512) \\ \midrule
\multirow{6}{*}{SPNews} & \multirow{3}{*}{LSTM} & & \multicolumn{1}{c}{\textbf{LN}} & \multicolumn{1}{c}{\textbf{LP}} & & \multicolumn{1}{c}{\textbf{LN}} & \multicolumn{1}{c}{\textbf{LP}} & & \multicolumn{1}{c}{\textbf{LN}} & \multicolumn{1}{c}{\textbf{LP}} \\ 
 & & \textbf{N+} & \multicolumn{1}{r}{1592/2379 (0.6692)} & 1511/2433 (0.6210) & \textbf{N+} & \multicolumn{1}{r}{1504/2461 (0.6111)} & 1076/1906 (0.5645) & \textbf{N+} & \multicolumn{1}{r}{1678/2667 (0.6291)} & 648/1106 (0.5858) \\
 & & \textbf{N-} & \multicolumn{1}{r}{2124/2537 (0.8372)} & 2359/2835 (0.8321) & \textbf{N-} & \multicolumn{1}{r}{2121/2610 (0.8126)} & 2626/3207 (0.8188) & \textbf{N-} & \multicolumn{1}{r}{2714/3445 (0.7956)} & 2354/2966 (0.7936) \\ \cmidrule(lr){2-11} 
 & \multirow{3}{*}{NGAT} & & \multicolumn{1}{c}{\textbf{LN}} & \multicolumn{1}{c}{\textbf{LP}} & & \multicolumn{1}{c}{\textbf{LN}} & \multicolumn{1}{c}{\textbf{LP}} & & \multicolumn{1}{c}{\textbf{LN}} & \multicolumn{1}{c}{\textbf{LP}} \\
 & & \textbf{N+} & \multicolumn{1}{r}{1734/2379 (0.7288)} & 1648/2433 (0.6773) & \textbf{N+} & \multicolumn{1}{r}{1694/2461 (0.6888)} & 1269/1906 (0.6657) & \textbf{N+} & \multicolumn{1}{r}{1844/2667 (0.6914)} & 624/1106 (0.5804) \\
 & & \textbf{N-} & \multicolumn{1}{r}{2040/2537 (0.8040)} & 2220/2835 (0.7830) & \textbf{N-} & \multicolumn{1}{r}{2035/2610 (0.7796)} & 2522/3207 (0.7864) & \textbf{N-} & \multicolumn{1}{r}{2667/3445 (0.7741)} & 2380/2966 (0.8024) \\ \bottomrule
\end{tabular}}
\end{table*}

\begin{table*}[]
\caption{Return Std Forecasting Performance (Five-run average). Notations: \textit{std\textsubscript{X}} denotes T=X.}
\label{tab:std}
\resizebox{\textwidth}{!}{%
\footnotesize 
\setlength{\tabcolsep}{3.5pt} 
\renewcommand{\arraystretch}{0.85} 
\begin{tabular}{lrrrrrrrrrrrr}
\toprule
\multirow{3}{*}{\textbf{Model}} & \multicolumn{6}{c}{\textbf{ACL 2018}} & \multicolumn{6}{c}{\textbf{SP News}} \\ \cmidrule(lr){2-7} \cmidrule(lr){8-13}
& \multicolumn{2}{c}{\textbf{std5}} & \multicolumn{2}{c}{\textbf{std10}} & \multicolumn{2}{c}{\textbf{std21}} & \multicolumn{2}{c}{\textbf{std5}} & \multicolumn{2}{c}{\textbf{std10}} & \multicolumn{2}{c}{\textbf{std21}} \\ \cmidrule(lr){2-3} \cmidrule(lr){4-5} \cmidrule(lr){6-7} \cmidrule(lr){8-9} \cmidrule(lr){10-11} \cmidrule(lr){12-13}
& R\textasciicircum{}2 & MSE & R\textasciicircum{}2 & MSE & R\textasciicircum{}2 & MSE & R\textasciicircum{}2 & MSE & R\textasciicircum{}2 & MSE & R\textasciicircum{}2 & MSE \\ \midrule
LSTM       & 0.04 & 0.39 & 0.26 & 0.24 & 0.35 & 0.22 & 0.16 & 0.66 & 0.27 & 0.44 & 0.36 & 0.31 \\
GCN        & 0.05 & 0.38 & 0.27 & 0.23 & 0.39 & 0.20 & 0.15 & 0.66 & 0.26 & 0.45 & 0.33 & 0.33 \\
GAT        & 0.06 & 0.38 & 0.30 & 0.22 & 0.40 & 0.19 & 0.13 & 0.67 & 0.24 & 0.46 & 0.33 & 0.33 \\
LSTM+GCN   & 0.03 & 0.39 & 0.26 & 0.24 & 0.36 & 0.21 & 0.15 & 0.66 & 0.26 & 0.44 & 0.34 & 0.32 \\
TGC        & 0.01 & 0.40 & 0.23 & 0.25 & 0.35 & 0.22 & 0.14 & 0.68 & 0.24 & 0.47 & 0.28 & 0.36 \\
AD-GAT     & -0.04 & 0.42 & 0.20 & 0.25 & 0.36 & 0.19 & 0.04 & 0.75 & 0.15 & 0.52 & 0.23 & 0.38 \\
NGAT (Our) & \textbf{0.07} & \textbf{0.38} & \textbf{0.31} & \textbf{0.22} & \textbf{0.45} & \textbf{0.17} & \textbf{0.19} & \textbf{0.63} & \textbf{0.31} & \textbf{0.42} & \textbf{0.38} & \textbf{0.30} \\ 
\bottomrule
\end{tabular}%
}
\end{table*}

\begin{table*}[]
\caption{Performances comparison of different graph construction methods on SPNews Dataset. Number in parathesis means memory window $\delta$}
\label{tab:graph-construction-compare}
\resizebox{\textwidth}{!}{
\begin{tabular}{lrrrrrrrrrrrr}
\toprule
\multirow{3}{*}{\textbf{\begin{tabular}[c]{@{}l@{}}Graph Construction\\ Method\end{tabular}}} & \multicolumn{6}{c}{\textbf{GAT}} & \multicolumn{6}{c}{\textbf{NGAT}} \\ \cmidrule(lr){2-7} \cmidrule(lr){8-13}
& \multicolumn{2}{c}{\textbf{std5}} & \multicolumn{2}{c}{\textbf{std10}} & \multicolumn{2}{c}{\textbf{std21}} & \multicolumn{2}{c}{\textbf{std5}} & \multicolumn{2}{c}{\textbf{std10}} & \multicolumn{2}{c}{\textbf{std21}} \\ \cmidrule(lr){2-3} \cmidrule(lr){4-5}\cmidrule(lr){6-7} \cmidrule(lr){8-9} \cmidrule(lr){10-11} \cmidrule(lr){12-13}
& \multicolumn{1}{c}{R\textasciicircum{}2} & \multicolumn{1}{c}{MSE} & \multicolumn{1}{c}{R\textasciicircum{}2} & \multicolumn{1}{c}{MSE} & \multicolumn{1}{c}{R\textasciicircum{}2} & \multicolumn{1}{c}{MSE} & \multicolumn{1}{c}{R\textasciicircum{}2} & \multicolumn{1}{c}{MSE} & \multicolumn{1}{c}{R\textasciicircum{}2} & \multicolumn{1}{c}{MSE} & \multicolumn{1}{c}{R\textasciicircum{}2} & \multicolumn{1}{c}{MSE} \\ \midrule
Static Graph & 0.1383 & 0.6799 & 0.2453 & 0.4616 & 0.3350 & 0.3334 & 0.1923 & 0.6448 & 0.3102 & 0.4258 & 0.3746 & 0.3150 \\ \midrule
Correlation (5 day) & 0.1008 & 0.7081 & 0.1878 & 0.4955 & 0.2645 & 0.3668 & 0.1836 & 0.6514 & 0.3014 & 0.4317 & 0.3607 & 0.3221 \\
Cooccurence (5 day) & \textbf{0.1550} & \textbf{0.6703} & \textbf{0.2704} & \textbf{0.4492} & \textbf{0.3505} & \textbf{0.3271} & \textbf{0.1999} & \textbf{0.6382} & \textbf{0.3128} & 0.4239 & \textbf{0.3771} & \textbf{0.3132} \\ \midrule
Correlation (21 day) & 0.0603 & 0.7434 & 0.1133 & 0.5402 & 0.1773 & 0.4113 & 0.1869 & 0.6491 & 0.3069 & \textbf{0.4181} & 0.3695 & 0.3170 \\
Cooccurence (21 day) & 0.1175 & 0.6944 & 0.2109 & 0.4814 & 0.2856 & 0.3579 & 0.1872 & 0.6489 & 0.2991 & 0.4332 & 0.3591 & 0.3218 \\ \midrule
\end{tabular}%
}
\end{table*}

\textbf{Volatility Analysis} As shown in Table~\ref{tab:std}, NGAT consistently outperforms all baselines in volatility forecasting across both datasets. For ACL2018, graph-based models demonstrate clear advantages over sequential models, validating the importance of relational information. On SPNews, while text-derived graphs show limited improvements due to potential noise in news-to-graph translation, NGAT maintains robust performance, confirming its ability to extract reliable signals even from imperfect relational data. The superiority of NGAT over other graph architectures highlights the effectiveness of its node-level attention mechanisms in interpreting financial relationships. Notably, monthly volatility prediction (T=21) achieves the highest performance across all three settings. This superiority aligns with financial practicality: monthly volatility (T=21) holds broader portfolio applications than noisy short-term measures (T=5/10). Combining the results from Table \ref{tab:r-mean} and Table \ref{tab:std}, aside from the NGAT model, no graph-based model (including both generic graph architectures and finance-specific SOTA variants) has demonstrated comprehensive superiority over the standalone LSTM sequence model across all forecast horizons in both tasks. This phenomenon arises because different prediction horizons require capturing distinct momentum spillover effects. For instance, highly connected companies (those with numerous edges) may experience varying influential counterparts across forecast horizons, whereas sparsely connected entities might show minimal horizon-dependent variations. This renders shared feature aggregation mechanisms inefficient, ultimately proving the limitations of inter-node shared feature aggregation while validating the effectiveness of node-level attention mechanisms in adapting to temporal dynamics. It is also noteworthy that, as the forecast horizon $T$ increases, performance on return trend prediction slightly declines, while volatility prediction improves significantly. Thus, setting $T=21$ provides a balance between performance on both tasks. Investors can make decisions based on predicted return trends and volatility, tailored to their risk preferences over the selected forecast horizon.


\textbf{Graph Construction Methods}  We compare text-derived dynamic graphs (news/tweets co-occurrence) against conventional return correlation graphs and static baselines (Table~\ref{tab:graph-construction-compare}), revealing three findings: 1) NGAT consistently surpasses GAT across all graph types, demonstrating the architectural advantage of node-specific attention mechanisms;  
2) Memory length optimization proves critical, with 5-day windows maximizing performance – particularly for news-based graphs where outdated content diminishes market relevance;  
3) Downstream evaluation alone provides incomplete graph quality assessment: While GAT favours co-occurrence graphs over correlation-based ones (suggesting better graph quality), this distinction diminishes under NGAT due to its ability to mitigate graph construction imperfections – demonstrating how advanced architectures can obscure inherent graph deficiencies.

\section{Conclusion}
This paper proposes the Node-level Graph Attention Network (NGAT) for modelling corporate relationship graphs, along with novel stock return/volatility prediction tasks extensible to long-term horizons. Extensive experiments validate NGAT's performance advantages and the practical utility of long-term forecasting. Though NGAT incurs higher computational complexity, this remains acceptable under fixed asset universe settings common in institutional portfolios.

\begin{credits}
\subsubsection{\ackname} This publication has emanated from research conducted with the financial
support of Science Foundation Ireland under Grant number 18/CRT/6183. For the purpose
of Open Access, the author has applied a CC BY public copyright licence to any
Author Accepted Manuscript version arising from this submission.

\end{credits}

%
%
\bibliographystyle{splncs04}
\bibliography{reference}

\begin{thebibliography}{10}
\providecommand{\url}[1]{\texttt{#1}}
\providecommand{\urlprefix}{URL }
\providecommand{\doi}[1]{https://doi.org/#1}

\bibitem{atkins2018}
Atkins, A., Niranjan, M., Gerding, E.: Financial news predicts stock market volatility better than close price. The Journal of Finance and Data Science  \textbf{4}(2),  120--137 (2018)

\bibitem{bojchevski2017deep}
Bojchevski, A., G{\"u}nnemann, S.: Deep gaussian embedding of graphs: Unsupervised inductive learning via ranking. arXiv preprint arXiv:1707.03815  (2017)

\bibitem{bollen2011twitter}
Bollen, J., Mao, H., Zeng, X.: Twitter mood predicts the stock market. Journal of Computational Science  \textbf{2}(1), ~1--8 (2011)

\bibitem{carhart1997persistence}
Carhart, M.M.: On persistence in mutual fund performance. The Journal of finance  \textbf{52}(1),  57--82 (1997)

\bibitem{chen2018incorporating}
Chen, Y., Wei, Z., Huang, X.: Incorporating corporation relationship via graph convolutional neural networks for stock price prediction. In: Proceedings of the 27th ACM international conference on information and knowledge management. pp. 1655--1658 (2018)

\bibitem{cheng2021modeling}
Cheng, R., Li, Q.: Modeling the momentum spillover effect for stock prediction via attribute-driven graph attention networks. In: Proceedings of the AAAI Conference on artificial intelligence. vol.~35, pp. 55--62 (2021)

\bibitem{cho2014learning}
Cho, K., Van~Merri{\"e}nboer, B., Gulcehre, C., Bahdanau, D., Bougares, F., Schwenk, H., Bengio, Y.: Learning phrase representations using rnn encoder-decoder for statistical machine translation. arXiv preprint arXiv:1406.1078  (2014)

\bibitem{corsi2009simple}
Corsi, F.: A simple approximate long-memory model of realized volatility. Journal of Financial Econometrics  \textbf{7}(2),  174--196 (2009)

\bibitem{persio2021forward}
Di~Persio, L., Garbelli, M., Wallbaum, K.: Forward-looking volatility estimation for risk-managed investment strategies during the covid-19 crisis. Risks  \textbf{9}(2) (2021)

\bibitem{engle1993measuring}
Engle, R.F., Ng, V.K.: Measuring and testing the impact of news on volatility. The Journal of Finance  \textbf{48}(5),  1749--1778 (1993)

\bibitem{fama2015five}
Fama, E.F., French, K.R.: A five-factor asset pricing model. Journal of financial economics  \textbf{116}(1),  1--22 (2015)

\bibitem{feng2019temporal}
Feng, F., He, X., Wang, X., Luo, C., Liu, Y., Chua, T.S.: Temporal relational ranking for stock prediction. ACM Transactions on Information Systems (TOIS)  \textbf{37}(2),  1--30 (2019)

\bibitem{glorot2010understanding}
Glorot, X., Bengio, Y.: Understanding the difficulty of training deep feedforward neural networks. In: Proceedings of the thirteenth international conference on artificial intelligence and statistics. pp. 249--256. JMLR Workshop and Conference Proceedings (2010)

\bibitem{jin2020stock}
Jin, Z., Yang, Y., Liu, Y.: Stock closing price prediction based on sentiment analysis and lstm. Neural Computing and Applications  \textbf{32},  9713--9729 (2020)

\bibitem{kim2018forecasting}
Kim, H.Y., Won, C.H.: Forecasting the volatility of stock price index: A hybrid model integrating lstm with multiple garch-type models. Expert Systems with Applications  \textbf{103},  25--37 (2018)

\bibitem{kingma2014adam}
Kingma, D.P., Ba, J.: Adam: A method for stochastic optimization. arXiv preprint arXiv:1412.6980  (2014)

\bibitem{kipf2016semi}
Kipf, T.N., Welling, M.: Semi-supervised classification with graph convolutional networks. arXiv preprint arXiv:1609.02907  (2016)

\bibitem{li2016tensor}
Li, Q., Chen, Y., Jiang, L.L., Li, P., Chen, H.: A tensor-based information framework for predicting the stock market. ACM Transactions on Information Systems (TOIS)  \textbf{34}(2),  1--30 (2016)

\bibitem{li2020multimodal}
Li, Q., Tan, J., Wang, J., Chen, H.: A multimodal event-driven lstm model for stock prediction using online news. IEEE Transactions on Knowledge and Data Engineering  \textbf{33}(10),  3323--3337 (2020)

\bibitem{li2021modeling}
Li, W., Bao, R., Harimoto, K., Chen, D., Xu, J., Su, Q.: Modeling the stock relation with graph network for overnight stock movement prediction. In: Proceedings of the twenty-ninth international conference on international joint conferences on artificial intelligence. pp. 4541--4547 (2021)

\bibitem{lo1990when}
Lo, A.W., MacKinlay, A.C.: When are contrarian profits due to stock market overreaction? The Review of Financial Studies  \textbf{3}(2),  175--205 (1990)

\bibitem{nelson1991conditional}
Nelson, D.B.: Conditional heteroskedasticity in asset returns: A new approach. Econometrica  \textbf{59}(2),  347--370 (1991)

\bibitem{niu2024evaluating}
Niu, Y., Lu, L., Dolphin, R., Poti, V., Dong, R.: Evaluating financial relational graphs: Interpretation before prediction. In: Proceedings of the 5th ACM International Conference on AI in Finance. pp. 564--572 (2024)

\bibitem{poterba1988mean}
Poterba, J.M., Summers, L.H.: Mean reversion in stock prices: Evidence and implications. Journal of financial economics  \textbf{22}(1),  27--59 (1988)

\bibitem{selvin2017stock}
Selvin, S., Vinayakumar, R., Gopalakrishnan, E., Menon, V.K., Soman, K.: Stock price prediction using lstm, rnn and cnn-sliding window model. In: 2017 international conference on advances in computing, communications and informatics (icacci). pp. 1643--1647. IEEE (2017)

\bibitem{soun2022accurate}
Soun, Y., Yoo, J., Cho, M., Jeon, J., Kang, U.: Accurate stock movement prediction with self-supervised learning from sparse noisy tweets. In: 2022 IEEE International Conference on Big Data (Big Data). pp. 1691--1700. IEEE (2022)

\bibitem{sunny2020deep}
Sunny, M.A.I., Maswood, M.M.S., Alharbi, A.G.: Deep learning-based stock price prediction using lstm and bi-directional lstm model. In: 2020 2nd novel intelligent and leading emerging sciences conference (NILES). pp. 87--92. IEEE (2020)

\bibitem{velivckovic2017graph}
Veli{\v{c}}kovi{\'c}, P., Cucurull, G., Casanova, A., Romero, A., Lio, P., Bengio, Y.: Graph attention networks. arXiv preprint arXiv:1710.10903  (2017)

\bibitem{xiang2022temporal}
Xiang, S., Cheng, D., Shang, C., Zhang, Y., Liang, Y.: Temporal and heterogeneous graph neural network for financial time series prediction. In: Proceedings of the 31st ACM international conference on information \& knowledge management. pp. 3584--3593 (2022)

\bibitem{xu2022hgnn}
Xu, C., Huang, H., Ying, X., Gao, J., Li, Z., Zhang, P., Xiao, J., Zhang, J., Luo, J.: Hgnn: Hierarchical graph neural network for predicting the classification of price-limit-hitting stocks. Information Sciences  \textbf{607},  783--798 (2022)

\bibitem{xu-cohen-2018-stock}
Xu, Y., Cohen, S.B.: Stock movement prediction from tweets and historical prices. In: Gurevych, I., Miyao, Y. (eds.) Proceedings of the 56th Annual Meeting of the Association for Computational Linguistics (Volume 1: Long Papers). pp. 1970--1979. Association for Computational Linguistics, Melbourne, Australia (2018)

\end{thebibliography}

%




\end{document}